\input{psfig.sty}

\documentstyle[11pt,aaspp4, flushrt]{article}  

\def\Rg{R_{\rm g}}
\def\taues{\tau_{\rm es}}
\def\kT{k T_{\rm e}}
\def\Ti{T_{\rm ion}}
\def\Rin{R_{\rm in}}
\def\LEdd{L_{\rm Edd}}
\def\MSun{M_{\odot}}
\def\rms{R_{\rm ms}}
\def\rtr{R_{\rm tr}}
\def\Ltot{L_{\rm tot}}
\def\Fhard{F_{\rm hard}}
\def\Fsoft{F_{\rm soft}}
\def\Lhard{L_{\rm hard}}
\def\Lsoft{L_{\rm soft}}

\def\LX{L_{\rm X}}
\def\Ls{L_{\rm s}}
\def\Lsb{L_{\rm s, blob}}
\def\Qphard{Q^{+}_{\rm hot}}
\def\Qp{Q^{+}}
\def\TIC{T_{\rm IC}}
\def\fadv{f_{\rm adv}}
\def\fcor{f_{\rm cor}}
\def\mus{\mu_{\rm s}}
\def\Ptrans{P_{\rm trans}}
\def\Prefl{P_{\rm refl}}
\def\MEdd{\dot M_{\rm Edd}}
\def\d{{\rm d}}

\begin{document}
 
\title{Probing the Inner Region of Cyg X--1 in the Low/Hard 
State through its X-ray Broad Band Spectrum}

\author{T. Di Salvo\altaffilmark{1}, 
C. Done\altaffilmark{2},
P. T. \.{Z}ycki\altaffilmark{3},
L. Burderi\altaffilmark{4},
N. R. Robba\altaffilmark{1}}
\altaffiltext{1}{Dipartimento di Scienze Fisiche ed Astronomiche, 
Universit\`a di Palermo, via Archirafi n.36, 90123 Palermo, Italy}
\altaffiltext{2}{Department of Physics, University of Durham, South Road, 
Durham, UK}
\altaffiltext{3}{Nicolaus Copernicus Astronomical Center, Bartycka 18,
00-716 Warsaw, Poland}
\altaffiltext{4}{Osservatorio Astronomico di Roma, Via Frascati 33, 
00040 Monteporzio Catone (Roma), Italy}
  
\authoremail{disalvo@gifco.fisica.unipa.it}

\lefthead{Probing the Inner Region of Cyg X--1}

\begin{abstract}

We present the broad band X-ray spectrum of Cyg X--1 in the low/hard
state as observed by the instruments on board of BeppoSAX. The spectrum 
spans from 0.1 to 200~keV, allowing the total accretion
luminosity to be observed, rather than extrapolated,
corresponding to $\sim 2$ per cent of the Eddington limit for a
10 $\MSun$ black hole.
The broad bandpass allows us to determine the continuum shape with great 
accuracy. Simple models of Compton up-scattering of seed photons
from the accretion disk do not adequately match the spectrum. 
At low energies an additional continuum component is required,
giving a complex soft excess which extends up to $\sim 4$ keV,
in line with previous results from ASCA. Moreover we
clearly detect a reflected component from the accretion disk which is
smeared, probably because of relativistic and Doppler effects. 
The reflecting material is not strongly ionized
and does not subtend a large solid angle as seen from the corona ($\Omega/
2\pi \sim 0.1-0.3$). The inner radius of the disk, that depends on the 
inclination of the system, is most probably between 10 and 70 
gravitational radii ($\Rg$).
An unsmeared reprocessed component, probably 
originating from the companion star or the outer disk, could also be present. 
In this case, the inner radius of the disk, as inferred from the smeared 
reflection, is smaller, between 6 and 20 $\Rg$. 
\end{abstract}


\section{Introduction}

Cyg X--1 is the brightest of the persistent galactic black hole
candidates.  It is a binary system consisting of a black hole
accreting matter from a supergiant type O star.  It shows (at least)
two different spectral states, namely a high/soft state where the
luminosity is dominated by strong blackbody emission at low energies
($kT = 0.5-1$ keV), and a steep power law (photon index $\Gamma > 2-3$), 
extending out to at least 800 keV, is also present, and a low/hard
state, which has much less soft emission ($kT = 0.1-0.3$ keV), and
has the luminosity dominated by a hard power law 
($\Gamma \sim 1.4-1.9$) which cuts off at $\sim 200$ keV.  In addition
to the continuum emission other spectral features are present in the
spectrum, most noticeably an iron K$_\alpha$ line and edge at 6.4
and 7.1 keV respectively.  These features are seen in both
spectral states, and can be fit with models of X-ray reflection of
the power law from an accretion disk (Done et al.\ 1992; 
Gierli\'{n}ski et al.\ 1997, 1999).

Detailed spectroscopy of the reflected spectrum can be used to probe
the accretion flow, since its amplitude and shape are
determined by the ionization state, geometry and velocity field of the
reprocessing plasma. Ionization of the reprocessor decreases the
photoabsorption opacity, thus increasing the reflection albedo at low
energies, as well as the energy and intensity of the iron
K$_\alpha$ fluorescence/recombination line, and energy and depth of
the iron K edge (Lightman \& White 1988; Ross \& Fabian 1993; 
\.{Z}ycki et al.\ 1994). The
normalization of the reflected spectrum for a given ionization then
depends on the inclination angle of the accretion disk with respect to
the observer and on the solid angle subtended by the disk to the hard
X-ray source (George \& Fabian 1991; Matt, Perola \& Piro 1991). The
combination of Doppler effects from the high orbital velocities and
strong gravity in the vicinity of a black hole smears the reflected
spectrum, so that the line (and the reflected continuum) has a
characteristically skewed, broad profile, whose detailed shape 
depends on the inclination and on how deep the accretion disk extends into
the black hole potential (Fabian et al.\ 1989).

The best known example of the effects mentioned above is the Fe K$_\alpha$ line
profile in the spectrum of the Seyfert 1 galaxy MCG--6--30--15 (Tanaka et
al.\ 1995). The amount of reflected spectrum is that expected for a disk
which covers half the sky as seen from the X-ray source (Lee et al.\ 1999). 
The fit of
the iron line profile implies that the inner edge of this reprocessor lies
at (or even within) the marginally stable orbit in Schwarzschild metric 
$\rms = 6\,\Rg$ ($\Rg\equiv G M/c^2$, where $G$ is the gravitational constant,
$M$ the mass of the black hole, and $c$ the speed of light) (Tanaka et al.\  
1995; Iwasawa et al.\ 1996). These effects should also be present in
Galactic Black Hole Candidates (GBHC), and should be more easily 
observable because of the much higher
signal-to-noise ratio. Yet in the low/hard state of Cyg X--1 the amount
of observable reflection is smaller than expected for a
complete disk (e.g Gierli\'{n}ski et al. 1997) and the
amount of relativistic smearing is less than that seen in
MCG--6--30--15, implying that the disk terminates at $10-20\Rg$ rather
than extending down to the last stable orbit (Done \& \.{Z}ycki
1999). Other low/hard state data from GBHCs show 
the same lack of
extreme relativistic smearing as Cyg X--1 (\.{Z}ycki, Done \& Smith 1997;
1998; 1999; Gilfanov, Churazov, \& Revnivtsev 1999),
as do some AGN (IC 4329a: Done, Madejski, \& \.{Z}ycki 2000;
NGC 5548: Chiang et al. 2000). In this respect MCG--6--30--15 has an 
exceptionally steep X-ray spectrum, so it is more comparable to the high/soft
state of GBHCs rather than to the low/hard state discussed here (Zdziarski, 
Lubi\'{n}ski \& Smith 1999).

If the 'standard' optically thick, geometrically thin disk does
truncate before the last stable orbit this could indicate that
inside this transition radius ($\rtr > \rms$) the accretion
changes to a hot, optically thin, geometrically thick flow, 
most plausibly an advection dominated accretion flow  (ADAF: 
see e.g.\ Narayan 1997 for a review). Such a geometrical 
configuration is compatible with the overall broad band spectral shape
of Cyg X--1, which has a rather hard (photon-starved) spectrum, as well
as with the observed low amplitude of reflection and relativistic
smearing (Poutanen, Krolik \& Ryde 1997; Gierli\'{n}ski et al.\ 1997; Dove et
al.\ 1997; review in Poutanen 1998). The alternative models which keep
an untruncated disk are reviewed in Done \& \.{Z}ycki (1999).  These models 
require that the hard X-rays are produced in active regions above the
disk, probably powered by magnetic reconnection (Galeev, Rosner \&
Vaiana 1979; Haardt, Maraschi \& Ghisellini 1994).  There are two 
plausible ways for the untruncated-disk models to match the spectral
constraints. First, the models in which the X-ray emission regions
expand with relativistic velocities away from the disk (Beloborodov
1999a,b):  this suppresses both the seed photons for Compton scattering
(so giving the observed hard spectrum) and the amount of reflection (by
beaming the observed X-ray spectrum away from the disk).
The second possibility is that the
X-ray illumination photo-ionizes the disk reducing the 
photo-electric opacity.  In this case there are no spectral features in the
reflected spectrum, since it is formed purely from Thomson
scattering. The large fraction of flux reflected means that this also
decreases the thermalized fraction of the primary X-rays, so there are
few soft photons (Ross, Fabian \& Young 1999; Nayakshin, Kazanas \& Kallman
2000).

Distinguishing between these various models should surely be possible
from detailed studies of the reflected spectrum and energetics of the
continuum spectrum. However, the derived parameters for the amount of
reflection and smearing depend crucially on how the continuum is
modelled and there is increasing evidence that simple thermal
disk blackbody and comptonized power laws are not sufficient to
describe the spectrum, especially at low energies where ASCA data show
an additional soft component below 3--4 keV in Cyg X--1 (Ebisawa et al.\ 
1996).  The continuum can be better characterized with a
broad bandpass, and we report here on the BeppoSAX 0.1--200 keV observation
of Cyg X--1 in the low/hard state. We show that there is indeed a
complex soft component,
and that the reflected continuum can be clearly disentangled from this
complex spectrum.

\section{Observations}

The BeppoSAX Narrow Field Instruments (NFIs) observed Cyg X--1 on 1998 May 3 
and 4, for an effective exposure 
time of 25 ks. The NFIs are four co-aligned instruments which cover more
than three decades of energy, from 0.1 keV up to 200 keV, with good
spectral resolution in the whole range. LECS (operating in the range 0.1--10
keV, Parmar et al.\ 1997) and MECS (1--11 keV, Boella et al.\ 1997) have
imaging capabilities with a Field of View (FOV) of $20'$ and $30'$ radius 
respectively. In these FOVs we selected the data for the scientific analysis 
in a circular region of $8'$ and $4'$ radius for LECS and MECS, 
respectively, around the centroid of the source. The background subtraction
was obtained using blank sky observations in which we extracted the 
background data in a region of the FOV similar to that used for the
source. HPGSPC (7--60 keV, Manzo et al.\ 1997) and PDS (13--200 keV, Frontera
et al.\ 1997) do not have imaging capabilities, because the FOVs, of $\sim 
1^\circ$ FWHM, are delimited by collimators. The background subtraction
for these instruments was obtained using the off-source data accumulated 
during the rocking of the collimators.
The energy ranges used for each NFI are: 0.12--3 keV for the LECS, 1.8--10 keV
for the MECS, 8--30 keV for the HPGSPC and 20--200 keV for the PDS. We added
a (conservative) systematic error of 5\% to the PDS data, 
to take into account calibration residuals\footnote{see the web page at
http://www.sdc.asi.it/software/cookbook/matrices.html}. Different 
normalizations of the four NFIs are considered by including in the model
normalizing factors, fixed to 1 for the MECS and kept free for the other
instruments. The same method has been used to fit Crab spectra 
(Massaro et al. 2000), obtaining photon indices
which are within $\Delta\alpha \sim 0.03$ of those expected from models 
of the synchrotron emission (Aharonian \& Atoyan 1995).

During the BeppoSAX observation Cyg X--1 was in its usual low/hard state
with a total (0.1--200 keV) unabsorbed luminosity of $\sim 3 \times 
10^{37}$ ergs/s, adopting a distance of 2 kpc (e.g.\ 
Massey, Johnson, \& Degioia-Eastwood 1995, Malysheva 1997).
Figure 1 shows the MECS light curves in three energy bands, 1--4 keV (first
panel from above), 4--7 keV (second panel) and 7--11 keV (third panel), and 
the corresponding hardness ratios. 
Although the MECS light curves show an increase of the
intensity up to $\sim 30\;\%$ during the observation, the hardness ratios 
appear to be constant, except for a little hardening at the end of the
observing period (between $5.5 \times 10^4$ and $6 \times 10^4$ s). This 
hardening is probably due to an absorption dip, because it is visible in the 
soft range and disappears in the hard range. Cyg X--1 is known to show
intensity dips, which preferentially occur near the superior conjunction,
at the orbital phase $\phi = 0$, although dips were also observed at
$\phi = 0.88$ and $\phi = 0.42$ (Ebisawa et al.\ 1996). 
The BeppoSAX observation spans the orbital phase interval $\phi = 0.39 - 
0.58$ (using the ephemeris reported in Gies \& Bolton 1982), 
and the observed dip occurs at phase $\phi = 0.51$.
We excluded this dip from the following spectral analysis.

\section{The Spectral Model}

The direct comptonized component from the corona is modeled by the
{\it thComp\/} model (Zdziarski, Johnson \& Magdziarz 1996), obtained by
solving the Kompaneets equation.

To describe the reflection component we use the angle dependent reflection 
model of Magdziarz \& Zdziarski (1995)
with a self-consistent iron emission line calculated for the given ionization 
state, temperature, spectral shape and metal abundances, as described in 
detail in \.{Z}ycki et al.\ (1999). The total reprocessed component can be
smeared to take into account the relativistic and kinematic effects of
disc emission (Fabian et al.\ 1989).  This is done by convolving a spectrum 
with the XSPEC {\it diskline\/} model, parametrized by the inner and 
outer radius of the disk, $R_{\rm in}$ and $R_{\rm out}$, and the irradiation
emissivity exponent, $\alpha$, $F_{\rm irr} \propto r^{-\alpha}$.  
We fix $R_{\rm out} = 10^4~ \Rg$, $\alpha=3$ and fit for $R_{\rm in}$, 
in order to obtain an estimate of the inner radius of the accretion disc.

The reflected spectrum also depends on the inclination of the system 
with respect to the line of sight, but
this parameter is more difficult to obtain for Cyg X--1 than for most
other GBHCs since the companion O star
loses matter via a strong wind, so it does not need to completely fill
its Roche lobe in order to provide the accreting material. The lack of
X-ray eclipses requires $i < 64^\circ$, and almost all angles between
this and $\sim 25^\circ$ have been derived from spectrophotometric
studies (see e.g.\ Ninkov, Walker \& Yang 1987). Recent Doppler
tomography studies of the H$\alpha$ and HeII lines set an upper limit
to the inclination of $55^\circ$ from the fact that the emission lines
from the stream were not eclipsed (Sowers et al.\ 1998). However, this
computation used the companion star radii of Gies and Bolton (1986), which have
been shown to be overestimated (Herrero et al.\ 1995). Thus we
repeat the fits for a wider range of inclination angles to quantify
how sensitive our results are to this parameter. Other unknown
parameters affecting the reflected spectrum are the elemental
abundances. We fix these at solar (Morrison \& McCammon 1983), except
for iron which is allowed to vary.

\section{Spectral Analysis}

Spectral analysis was performed in XSPEC v.\ 10 (Arnaud 1996), 
with all the non-standard spectral models implemented as local models.

In Figure 2 we show the result of a fit of the 1.8--30 keV spectrum 
of Cyg X--1 with a simple photoelectric absorbed power law (the $\chi^2/d.o.f.$
was 2008/274). In the residuals (in unit of $\sigma$, lower panel) all the 
features expected in Cyg X--1 spectrum are clearly visible, namely the soft 
excess, the iron line and edge, and the spectral hardening due to the 
reflection at $\sim$ 20 keV. 

We tried to fit the 0.1--200 keV energy spectrum of Cyg X--1 with a simple 
model consisting of a direct comptonized spectrum described by the 
{\it thComp\/}  model, the corresponding reflected component, a blackbody to
describe the soft emission, and photoelectric absorption by cold matter.
The temperature of the soft seed photons for the Comptonization was fixed
to the temperature of the blackbody.  We also fixed the iron abundance to the 
solar value and $\cos i = 0.6$, where $i$ is the inclination angle 
(see Done \&  \.{Z}ycki, 1999). 
This model gives a poor fit with a $\chi^2/d.o.f. = 934/681$. In the
fit residuals there are indications for a more complex shape of the
soft component.

\subsection{The soft component}

Ebisawa et al.\ (1996) studying ASCA data of Cyg X--1 in the hard
state, found that the shape of the soft component is not well
described by a simple blackbody spectrum. They obtained a good fit
using a blackbody plus a steep power-law tail. This is a rather
unphysical description as the bolometric luminosity is infinite, so
instead we use an additional comptonized spectrum described by the
{\it comptt\/} model (Titarchuk 1994).  This improved the fit
significantly, giving $\chi^2/d.o.f. = 713/679$, using a reflection
model without relativistic effects (model 1 in Table 1), and
$\chi^2/d.o.f. = 689/678$ considering the relativistic smearing of the
reflection component (model 2 in Table 1). In both cases the seed
photons for the additional comptonized component were tied to the
blackbody temperature, and the electron temperature is $\sim 3$ keV. 
The optical depths reported in Table 1 are calculated for spherical
geometry of the comptonizing region; for a slab geometry it is roughly 
half the previous value.
The large uncertainties on the temperature and optical depth of this 
component are due to the fact that these parameters are correlated to each 
other and are not well constrained individually.
We have also found that using a multi-color disk blackbody 
({\it diskbb} in XSPEC, Mitsuda et al.\ 1984) instead of a blackbody to fit 
the soft emission does not improve the fit.

The soft excess is significantly detected using just the LECS and MECS
data, showing that it is not an artifact of slight discrepancies in
cross-calibration between the instruments, amplified by the broad
bandpass covered by BeppoSAX. A blackbody spectrum,
together with a comptonized continuum and relativistically smeared
reflection, gives $\chi^2/d.o.f.=566/437$, while adding the {\it comptt\/}
component this reduces to $\chi^2/d.o.f.=510/434$.

Using an additional blackbody, with a temperature of $\sim 0.4$ keV, instead 
of {\it comptt}, as a description of the soft excess, we obtain 
$\chi^2/d.o.f. = 785/680$, without relativistic smearing of the reflection 
component, and $\chi^2/d.o.f. = 749/679$ considering the relativistic effects.
The first description, using the {\it comptt} model for the soft excess, thus 
appears to be somewhat better, but we will consider both these possibilities 
in the Discussion. 

Different descriptions of the soft excess give somewhat different estimates
of the bolometric (unabsorbed) luminosity. In the first model (blackbody plus 
{\it comptt}) the unabsorbed  flux in the two soft components is 
$\Fsoft \approx 2.1\times 10^{-8}\ {\rm erg\ cm^{-2}\ s^{-1}}$, while it is 
only $3.3\times 10^{-9}\ {\rm erg\ cm^{-2}\ s^{-1}}$ in the second model
(two blackbodies). Using multi-color disk blackbody and {\it comptt\/}
we obtain rather larger
$\Fsoft \approx 7\times 10^{-8}\ {\rm erg\ cm^{-2}\ s^{-1}}$.
The hard flux is the same in all models, 
$\Fhard \approx 3.8\times 10^{-8}\ {\rm erg\ cm^{-2}\ s^{-1}}$.

\subsection{The hard component}

Adopting now the blackbody plus {\it comptt\/} as description of the
soft component we proceed to investigate the hard component in the spectrum.

\subsubsection{The Comptonized continuum}

The comptonized continuum is well described by the {\it thComp\/} model.
The spectral slope derived from the data is $\Gamma \approx 1.67 $ and the electron
temperature is $\kT \approx 140$ keV (these values were obtained for parameters
of the reprocessing component $\cos i=0.78$ and Fe abundance of $2\times$Solar, 
see Section~\ref{sec:reproc}, but they are nearly independent of particular
values chosen). The {thComp\/} model is known to give rather inaccurate
temperature estimations for temperatures above $\sim 100$ keV (A. Zdziarski, 
private communication). We have therefore performed Monte Carlo simulations
in order to derive the physical value of $\kT$. Using the Monte Carlo code
described in Appendix A of \.{Z}ycki et al.\ (1999) we obtain 
$\kT\approx 90$ keV. The value of temperature is nearly independent of the 
plasma cloud geometry  assumed, contrary to the value of optical depth,
which we will discuss later in Section~\ref{sec:hot_depth}.

\subsubsection{The reprocessed component}
\label{sec:reproc}

The total model with the reprocessed component without relativistic smearing
yields $\chi^2/d.o.f. = 713/679$ (model 1 in Table 1).
Allowing now for relativistic smearing of the 
reflection component we obtain a better fit with $\chi^2/d.o.f. = 689/678$
(model 2 in Table 1).  An F-test demonstrates that the probability
of chance improvement of the fit is $2 \times 10^{-6}$. 
In this case, the inner radius of the disk, as inferred from the reflection 
component, is $\sim 130~\Rg$, and the reflection amplitude is 
$\Omega /2\pi \sim 0.3$, where $\Omega$ is the solid angle subtended by the
reflector as viewed from the corona.  

To constrain the iron abundance and the inclination angle $i$
of the system we fitted the data using different values of these two 
parameters. In particular we considered $\cos i = 0.87, 0.71, 0.60, 0.50$ and 
[Fe] = 3, 2, 1.5, 1 $\times$ Solar abundance (the grid of models for the 
Fe K$_\alpha$ line does not allow us to test Fe abundances $>3\times$Solar). 
The resulting values of $\chi^2$ and $R_{\rm in}$ are reported in Table 2. 
We note that the inferred inner radius of the disk is
strongly dependent on the inclination as well as iron abundance. Generally
we obtain larger radii in correspondence of high inclinations, because
in this case the Doppler shifts, due to the orbital velocity of the matter,
are stronger, giving substantial broadening. Nevertheless, we always obtain
radii smaller than 150 $\Rg$, with the most probable values lower than
70 $\Rg$. The best overall fit is obtained for $\cos i = 0.87$ and 
[Fe] = 2, corresponding to an inner radius of $\sim 10~\Rg$. However we
obtain an equivalently good fit for $\cos i = 0.78$ and 
[Fe] = 1.5, corresponding to an inner radius of $\sim 70~\Rg$.

To compare the results we plotted in Figure 3 the contributions to the
$\chi^2$ as a function of the energy in the iron line range for various 
models.  Model 1 of Table 1 with a not relativistically smeared reflection
is shown in the upper panel, the model corresponding to $\cos i = 0.78$
and [Fe] = 1.5 is shown in the top middle panel, and the model corresponding
to $\cos i = 0.87$ and [Fe] = 2 is shown in the bottom middle panel.
We note that while the model with $\cos i = 0.87$ and [Fe] = 2
can better fit the region around 5 keV, it leaves some residuals between
5.5 and 6.5 keV. On the other hand the model with $\cos i = 0.78$ and 
[Fe] = 1.5 gives a better fit of the region between 5.5 and 6.5 keV.
Because the latter seems to be better in the iron line region, this
indicates that the inner radius of the disk is $\sim 70~\Rg$, or that a
more complex model is needed to fit the data.

A narrow (not relativistically smeared) and not ionized reflection component 
is also expected from the companion star and/or the outer flared disk,
and it was actually observed in ASCA and BBXRT data (Ebisawa et al.\
1996, Done \& \.{Z}ycki 1999).
Adding to model 2 in Table 1 another reprocessed component, not ionized and 
not smeared, we obtain  $\chi^2/d.o.f. = 683/677$, i.e.\ 
a reduction of the $\chi^2$ by $\Delta \chi^2 = 6$ (model 3 in Table 1). 
An F-test shows that this improvement is 
significant at 99\% confidence level for one additional parameter.
Also in this case we fitted BeppoSAX data using different values of
$\cos i$ and iron abundance.  The resulting values of $\chi^2$ and inner
radius are reported in Table 3. In this case we obtain much smaller
inner radii, around 10--20 $\Rg$, sometimes as small as 6 $\Rg$, {\it i.e.} the
last stable orbit in the Schwarzschild metric. The best fit has 
$\cos i = 0.78$ and [Fe] = 2, corresponding to $\Rin \sim 6~\Rg$.
With these two parameters we obtain the results shown in 
Table 1 (model 4).  In Figure 4a we show the Cyg X--1 data fit using this 
model (upper panel) and the residuals (in unit of $\sigma$) with respect 
to the model (lower panel).  The contributions to the $\chi^2$ as a 
function of the energy in the iron line range are also shown in Figure 3
(lower panel).

The results are subtly different from those derived from the ASCA data
(Done \& \.{Z}ycki 1999), where the best fit was for somewhat higher
inclination ($\cos i = 0.75-0.50$ and [Fe] = 2), with 
correspondingly larger inner disk radii. We caution that at this level
small uncertainties in the response and cross-calibration become important.

A comparison of Tables 2 and 3 shows that the additional unsmeared component
does not give a significant 
improvement of the fit for all combinations of inclination and iron 
abundance considered, so we cannot say it is unambiguously detected in our 
data.

\section{Discussion}

We analyzed the broad band (0.1--200 keV) spectrum of Cyg X--1 in the
hard state observed by BeppoSAX.  The total spectrum predicted by
model 4 from Table 1 is plotted in Figure 4b.  The overall spectrum
consists of a complex soft component and a hard component. The former
can be described by a blackbody (or disk blackbody) 
at $k T_0\sim 0.1$ keV and an additional
component which can be described by Comptonization of soft photons
in a low temperature ($k T \sim 2$ keV) plasma with 
moderate optical depth ($\tau \sim 6$ for a spherical geometry). 
This is plausible thought not a unique description -- the additional 
component can be for instance a second blackbody with $k T \approx 0.4$ keV. 

Such a complex model for the soft excess in Cyg X--1 is
in agreement with the results of Ebisawa et al.\ (1996) who fitted
the soft component with a blackbody and a steep power-law tail.  
The presence of such additional comptonized excesses is not uncommon
in X-ray spectra of other accreting black holes. They were previously reported
in the very high and high states of Nova Muscae 1991 (\.{Z}ycki et al.\ 
1998), in the high state of GRS~1915+105 and GRO~J1655--40 (Coppi 1999; 
Zhang et al.\ 2000), in the high state of Cyg X--1 (Gierli\'{n}ski et al.\ 1999) 
and also in the Seyfert 1 galaxy NGC~5548 (Magdziarz et al.\ 1998). They suggest
existence of a third phase of accreting plasma, intermediate in 
properties between the cold, optically thick disk and the hot, optically
thin matter responsible for the hard Comptonization. The origin and location 
of such warm plasma would be different in the three different scenarios 
discussed below.

The hard component
can be modelled as comptonized emission from a hot ($\sim 100$ keV) corona,
with the corresponding reprocessed component
(Fe K$_{\alpha}$ line and edge, and the reflected continuum). 
The reprocessed component has a small amplitude, 
$\Omega/2\pi \sim 0.1-0.3$, is smeared, and is not highly ionized.
The best fit indicates high values of the iron 
abundance (2 times Solar) and moderate inclination angles 
($\cos i = 0.7-0.8$).
A second not ionized and not smeared reflection can also be
present, probably from the companion star and/or an outer flared disk.


Smearing of the reflected component is significantly detected in Cyg
X--1. Assuming that it is due to relativistic effects, this indicates the 
presence of optically thick material at small radii.  The value of the
inner radius of the disk is strongly dependent on the inclination
angle of the system and the iron abundance (and on details of the
modelling and calibration).  The model fits given in Table 3,
using two reflected components, 
give values between 6 and 20 $\Rg$. Without a second reflector the inner 
radius can be larger ($\sim 150$ $\Rg$ at most, see Tab. 2), 
but the fits are generally worse.
These results are consistent with the inner radius of the disk inferred 
from the rapid timing variability properties, in the hypothesis that the QPO
frequency detected in these systems is related to the innermost disk radius. 
In the case of Cyg X--1 in the low/hard state, $\Rin$ should be less than
$\sim 20~\Rg$ (depending on the mass of the black hole; 
Di Matteo \& Psaltis, 1999). 

Our broad-band spectral results are in agreement with previous 
studies of the low/hard state of Cyg X--1 
(Poutanen et al.\ 1997; Gierli\'{n}ski et al.\ 1997;
Dove et al.\ 1997), where it was demonstrated that the observed
spectra are incompatible with the model of an accretion disk with 
a static corona in a plane parallel geometry (Haardt \& Maraschi 1993). 
In these models the soft photons emitted
by the disk have to pass through the corona, resulting in a strong cooling.
Therefore the predicted spectra are too soft to match the observed hard 
spectrum from Cyg X--1.
Instead, they can be interpreted within two geometrical
scenarios: a hot inner disk with cold outer disk, and a disk with 
active regions. We will now discuss both geometries in some detail.

\subsection{The hot disk model}

In the hot disk model, the hard X-ray spectrum originates in an inner
accretion flow, with electron temperature $\sim 100$ keV and optical
depth $\taues \sim 1$. If the hot flow is heated by a viscous ($\alpha P$)
mechanism (Shakura \& Sunyaev 1973), the electron temperature has to
be different than the ion temperature (Shapiro, Lightman \& Eardley
1976), in which case $\Ti$ approaches the virial temperature, the flow
becomes geometrically thick, and advective transfer of energy is
important (Ichimaru 1977, Narayan \& Yi 1995). 
Conduction of heat between the hot flow
and any `standard' cold disk necessarily leads to 
evaporation of the disk if the accretion rate
is smaller than a critical value (R\'{o}\.{z}a\'{n}ska \& Czerny
2000). This gives a mechanism for a transition between disk dominated
and hot flow dominated accretion, though there can be a region of
overlap between the two flows (R\'{o}\.{z}a\'{n}ska \& Czerny
2000), where the observed additional soft component may be generated. 

\subsubsection{Optical depth of the central plasma cloud}
\label{sec:hot_depth}

We have performed Monte Carlo simulations of Comptonization in this
geometry to estimate the optical depth of the central plasma cloud,
and the amplitude of the reprocessed component. Approximating the
complex geometry to a central, uniform density sphere and an outer
cold disk, without any overlap, we find that the observed hard
spectrum can be matched if $\kT\approx 90$ keV and $\taues\approx
1.8$. This then predicts the amplitude of the spectrum of the soft seed 
photons and the amount of reflected spectrum from the accretion disk. 
Both these predicted values
are within the observed limits, if we use the {\it comptt}
model to describe the soft excess. Using two blackbodies for the soft 
emission, the observed ratio $\Lhard/\Lsoft$ is larger than the predicted 
value in the above geometry, without overlap between the outer disk and 
the central plasma cloud.  We also note that the assumed
uniform sphere may not be a good approximation to a real accretion
flow, where energy dissipation is concentrated towards the center.

\subsubsection{The overall energetics}
\label{sec:budg}

Adopting the Comptonization model for the soft excess, 
we find the total flux in the hard component  
$\Fhard = 3.8 \times 10^{-8}\ {\rm erg\ cm^{-2}\ s^{-1}}$ and
in the soft component (blackbody plus the soft comptonized component) 
$\Fsoft = 2.1 \times 10^{-8}\ {\rm erg\ cm^{-2}\ s^{-1}}$.
The bolometric luminosity of the primary source is then
\begin{equation}
\label{equ:tot}
\Ltot = \Lhard + \Lsoft =  4\pi d^2 \Fhard + {2 \pi d^2 \over \cos i}
\Fsoft,
\end{equation}
assuming isotropic emission of the hard X-ray source, and disk emission
of the soft X-ray source. 
This gives 
$\Ltot = 2.5\times 10^{37}\, d_2^2$ erg s$^{-1}  \approx 0.02 
\,d_2^2 M_{10}^{-1} \LEdd$ 
for $\cos i=0.7$, where $d_2$ is the distance to Cyg X--1 in units of 2
kpc, $M_{10}$ is the black hole mass in units of $10 \MSun$, and
$\LEdd$ is the Eddington luminosity for a $10 \MSun$ black hole.

To compute the mass accretion rate we need to take into account that a fraction
of energy may be advected into the black hole. Denoting the advected 
fraction by $\fadv$ we obtain the total rate of viscous energy
dissipation,
$\Qp = \Lhard/(1-\fadv) + \Lsoft = 
\Ltot\{1 +(\Lhard/\Ltot)[\fadv/(1-\fadv)]\}$. The mass accretion rate is
then
\begin{equation}
\dot M = \Qp / (\eta c^2) \approx 4.9\times 10^{17}\, d_2^2 \,
\left(1 + {\Lhard \over \Ltot}{ \fadv \over 1-\fadv} \right)\ {\rm g\
s^{-1}}, 
\end{equation}
where $\eta = 0.057$ is the efficiency of accretion in Schwarzschild
metric.
This gives
\begin{equation}
\dot m \equiv {\dot M \over \MEdd}  \approx 0.02\, d_2^2 \, M_{10}^{-1}
\left(1 + {\Lhard \over \Ltot}{ \fadv \over 1-\fadv} \right),
\end{equation}
for $\MEdd = \LEdd/(\eta c^2) \approx 2.5\times 10^{19} M_{10}\ 
{\rm g\ s^{-1}}$. For example, for $\fadv=0.75$ (as in the solution of 
Zdziarski, 1998, for $\dot m = \dot m_{\rm crit}$), 
$\dot m \approx 0.06 \, d_2^2\, M_{10}$.

\subsubsection{Estimating the transition radius}

The transition radius does not enter the Monte Carlo simulations performed 
in Sec.~\ref{sec:hot_depth}: the 
fraction of soft photons intercepted by the hard source is completely
determined by the geometry, which is scale invariant. 
We can estimate the truncation radius using information from 
total energy budget (Section~\ref{sec:budg}).
This  again requires the assumption that there is relatively 
little overlap between the two phases, and the cold disk emission is 
predominantly
due to viscous energy dissipation rather than thermalization of the
illuminating X-rays. 

From Eq. (\ref{equ:tot}) the fraction of viscous dissipation taking place 
in the hot flow is
$\Qphard /\Qp \approx 2/(3-\fadv)$. 
The radial distribution of energy dissipation per unit area
of the disk is described by (e.g.\ Frank, King \& Raine 1985)
\begin{equation}
\label{equ:diss}
F(R) = {3 \over 8\pi} {G M \dot M \over R^3} \left(1-\sqrt{{6\Rg \over R}}
\right).
\end{equation}
Using Eq.~\ref{equ:diss} we can estimate $\rtr$ solving
\begin{equation}
{\Qphard \over \Qp } = { \int_{6\Rg}^{\rtr} 2\pi F(R) R dR \over 
            \int_{6\Rg}^{\infty} 2\pi F(R) R dR }.
\end{equation}
For a purely radiative inner flow ($\fadv=0$) we obtain $\rtr \approx 40
\,\Rg$. The  presence
of advection increases this estimate and for e.g.\ $\fadv=0.75$
we obtain $\rtr \approx 130\,\Rg$.

Alternatively, the inner radius of the cold disk can be estimated using the
information on the blackbody soft component, assuming that this
represents the emission from the inner part of a standard, 
optically thick, geometrically thin accretion disk. Its
luminosity is then the total potential energy that has been released at 
R$_{\rm in}$:
\begin{equation}
L_{\rm BB} \simeq \frac{G M \dot{M}}{2 R_{\rm in}}
\end{equation}
We attribute the measured temperature of the blackbody to the maximum
temperature in the disk, that is reached at the inner radius $R_{\rm in}$
where the disk is truncated:
\begin{equation} 
T_{\rm max} = \left(\frac{3 G M \dot{M}}{8 \pi \sigma R_{\rm in}^3} 
\right)^{1/4} \left(1 - \sqrt{\frac{6 \Rg}{R_{\rm in}}} \right)^{1/4}
\end{equation}
From these two equations we can find the inner radius of the disk as a 
function of the measured temperature and luminosity of the blackbody.
Using the values reported in Table 1 (model 4), and considering that 
$L_{\rm disk} = 2 \pi d^2 F_{\rm earth}/\cos i$, where $F_{\rm earth}$ 
is the observed flux, we obtain 
$R_{\rm in} \sqrt{\cos i} \simeq 35$ $\Rg$, for a distance of 2 kpc. 

This should be modified as a simple
blackbody spectrum does not adequately
describe the emission of accretion disks in X-ray binaries, because 
electron scattering modifies the spectrum (Shakura \& Sunyaev 1973; 
White, Stella \& Parmar 1988). In this case, the measured color temperature
is related to the effective temperature of the inner disk 
$T_{\rm col} = f_{\rm col} T_{\rm eff}$, where $f_{\rm col}$ is the spectral 
hardening factor.
The factor $f_{\rm col}$ has been estimated by Shimura \& Takahara (1995) to be
$\sim 1.7$ for a luminosity of $\sim 10\%$ of the Eddington limit, with a
little dependence on the mass of the compact object and the radial position.
Applying this correction to the values of $R_{\rm in}$ obtained above, we find 
$R_{\rm eff} \sqrt{\cos i} = f_{\rm col}^2 R_{\rm in} \sqrt{\cos i} \sim
100$ $\Rg$. The correction factor used here is probably an underestimate
since Shimura \& Takahara (1995) do not consider illuminated disks,
as it is in the case of Cyg X--1.
For instance the additional soft excess component 
could be the emission of the inner part of the disk, strongly modified
by the Comptonization, which would then imply $f_{\rm col}\sim 4-5$. However
we are only considering here the part of the disk (probably at large radii) 
that emits the blackbody spectrum. Therefore this region is probably not 
dramatically influenced by the Comptonization and a factor 1.7 could be good 
enough for this rough estimate.

\subsubsection{Summary of hot disk model}

The geometry with a central, uniform density, hot plasma cloud with no
overlap between this and an outer disk can reproduce the overall
observed emission. Energy balance and the observed soft seed
photon temperature both point to a transition radius between the
disk and the hot flow of $\sim 100~\Rg$, in agreement with the 
estimation of the inner disk radius from the observed
smearing of the iron line (assuming that this arises from relativistic
effects), when only one reflection component is considered (see Tab. 2). 
However, this is in conflict with the smearing of the iron line in the 
best fit models with two reflection components. 
For this model to fit the observations requires either that there 
is further smearing of the reflected component by
Comptonization (see section 5.2.2) and/or consideration of a more 
realistic geometry,  with a hot flow which is centrally
concentrated, and where there is some overlap between the hot and cold
flows.

\subsection{The active regions model}

Another possibility for explaining hard spectra such as those observed from
Cyg X--1 and other GBHCs is a geometry of an untruncated, cold disk
with clumpy active regions (Stern et al.\ 1995). 
These can be physically envisioned
as magnetic flares (Galeev et al.\ 1979; Haardt et al.\ 1994). 
However, in order to decrease the amplitude of the reprocessed
component in such a configuration, one has to postulate that either
the comptonizing plasma is not static but it outflows at mildly
relativistic speed, $\beta\sim 0.3$ (Beloborodov 1999a,b), or
a hot ionized skin forms on top of the disk, increasing the X-ray albedo,
reducing both the amplitude of the observed reflection and the soft, 
thermalized flux (Ross et al.\ 1999; Nayakshin et al.\ 2000).

\subsubsection{The outflow model}

Using the formulae from Beloborodov (1999a,b) we can estimate the
outflow velocity, $\beta$ (in units of the speed of light), the flare
geometry (parameterized by $\mu_s$, the cosine of the angle between
the midpoint of the flare and its edge as seen from the disk)
and the fraction of energy dissipated
in the magnetic corona, $f_{cor}$, from the observed spectral index
$\Gamma\approx 1.63$, reflection amplitude $f\approx 0.2-0.3$, and
ratio of luminosities, $\Lsoft/\Lhard \sim 0.5$.

Eq.~(4) of Beloborodov (1999a) for the soft luminosity intercepted by the
blob, can be generalized to the case of $\fcor<1$ to give
\begin{equation}
 \Lsb = \int_{-1}^{-\mus}\,[(1-a)\LX(\mu) + (1-\fcor) \Ltot] \d \mu, 
\end{equation}
where $a\sim 0.3$ is the energy-integrated X-ray albedo.
$\Ltot$ is the total energy dissipation rate, 
$\LX = \fcor \Ltot$ is the blob (hard X-ray) luminosity, 
and we have assumed the same distribution of emission for both soft 
X-ray components (the blackbody/disk-blackbody and additional
soft X-ray  emission).
This would be roughly equivalent to assuming the covering fraction of
the active regions not much less than 1, so that the spatial distribution
of the disk effective temperature is roughly uniform.
The angular distribution of $\LX$ in the observer frame is given by
\begin{equation}
 \LX(\mu) = { \LX \over 2 \gamma^4 (1-\beta \mu)^3 }.
\end{equation}
Beloborodov (1999a) demonstrated that the soft luminosity in the comoving
frame is $\Ls^{\rm c} = \Lsb \gamma^2[1-\beta(1+\mus)/2]$.
The amplification factor $A\equiv \LX/\Ls^{\rm c}$ can then be
directly related to the hard X-ray slope 
$\Gamma \approx 2.33 (A-1)^{-\delta}$, where $\delta\approx 1/6$ for GBHCs
(Beloborodov 1999a, 1999b).
Next, the observed soft X-ray luminosity,
i.e.\ {\it not\/} intercepted by the blob is
\begin{equation}
 \Lsoft = \int_{-\mus}^{0}\,[(1-a)\LX(\mu) + (1-\fcor) \Ltot] \d \mu,
\end{equation}
while the observed hard X-ray luminosity is $\Lhard = \LX(\cos i)$, where
$i$ is the inclination of the system. The three observables,
$\Gamma$, $f$ and $\Lsoft/\Lhard$, enable estimation of $\beta$, $\mus$
and $\fcor$ (assuming the inclination is known).

First, the outflow velocity can be directly obtained from the observed
amplitude of reflection. For $f=0.2-0.3$ and $\cos i=0.75$ 
we obtain $\beta=0.3-0.4$. Assuming for concreteness $\beta=0.35$,
we find that $\fcor \approx 0.8$ and $\mus\approx 0.5$
reproduce both
$\Gamma\approx 1.6$ and $\Lsoft/\Lhard \approx 0.5$ in the spectral model 
containing the comptonized excess. For the other model of the soft component
(two blackbodies), which gives much smaller $\Lsoft/\Lhard \approx 0.1$,
the coronal dissipation fraction is $\fcor > 0.99$ while $\mus\approx 0.3$.

A  further development of the original model by Beloborodov (1999a) 
may also be necessary in the light of the results on relativistic smearing
of the reprocessed component. The outflow velocity has to be larger
in the inner disk region than in the outer one, if the reprocessed 
component does not show the extreme smearing, corresponding to inner disk
radius close to the last stable orbit at $6\Rg$. However, our results here
are somewhat ambiguous and, in particular, in the model with two reprocessed
components, the relativistic smearing does indeed  correspond to inner radius
close to $\rms$. The approximation with constant $\beta(r)$ may thus be
appropriate. 

\subsubsection{The hot skin model}

In this model, assuming the disk is radially uniform,  
the observed amplitude of cold reflection can be expressed as
\begin{equation}
f \sim { [\Ptrans(\tau)]^2 \over 1+\Prefl(\tau) },
\end{equation}
where $\tau$ is the optical depth of the hot skin, while $\Ptrans(\tau)$
and $\Prefl(\tau)$ are transmission and reflection probabilities {\it
in the hot skin} (i.e.\ the ratio to the incident flux of the flux
transmitted down through the hot skin to the cold material, and the
flux reflected out of the top of the hot skin, respectively).
Obviously, for a pure scattering layer $\Ptrans(\tau) + \Prefl(\tau) = 1$.
We obtained both probabilities from  Monte Carlo simulations in a slab
geometry. The observed amplitude, $f_{\rm rel} \approx 0.3$ corresponds to
$\Ptrans(\tau) \approx 0.64$ and $\tau\approx 0.7$.

The ratio of the soft thermalized flux to the primary hard X-ray flux is 
$(1-a)\Ptrans(\tau) / (1+\Prefl(\tau)) \approx 0.4$, 
close to that observed 
in the model with the additional soft excess component described as 
comptonized emission
(see Section~\ref{sec:budg}). 
Indeed one could imagine that this additional
component is the disk emission comptonized in the hot skin, as the photons
diffuse through it. 
The expected ratio of fluxes of the blackbody component, 
$F_{\rm bb}$, to the comptonized soft excess component, $F_{\rm sxe}$, 
is $F_{\rm bb}/F_{\rm sxe} \approx  \Ptrans/[A(1-\Ptrans)]$,
where $A$ is the Compton amplification.
For an optical depth of $0.7$,   $F_{\rm bb}/F_{\rm sxe}$
should thus be $\approx 2$ ($A-1\ll 1$ here), while from the best
fit model we infer $F_{\rm bb}/F_{\rm sxe} \approx 4$.
Also, for an optical depth of $\sim 0.7$ the spectral
shape of the additional comptonized component requires that the 
temperature of the hot skin is $k T_{\rm skin} \approx 12$ keV.
This is a little high compared to the mean
Compton temperature in the skin of $k\TIC\approx 5$ 
keV (Nayakshin et al. 2000). 

The diffusion of the reprocessed photons through the hot skin results in
smearing of iron spectral features due to Comptonization. We have modelled 
this by convolving the reprocessed spectrum
with the Comptonization Green's functions of Titarchuk (1994).
The smearing in the data is consistent with Comptonization alone
for $k T_{\rm skin}=12$ keV and
$\tau=0.7$: fitting only MECS data in the 4--10 keV band
we have found that Comptonization smearing gives $\chi^2/d.o.f. = 130/126$ 
while relativistic smearing gives $\chi^2/d.o.f. = 127/125$.
However, in the hot skin model, there is no reason
for the disk to be truncated, and so both smearing mechanisms would operate.
The relativistic smearing would be that expected from a disk
extending to $6\,\Rg$, although with emissivity going to zero at the
last stable orbit (Shakura \& Sunyaev 1973), 
$F_{\rm irr}(r) \propto r^{-3}\,(1-\sqrt{6\Rg/r})$. 
We have found that the model with both the relativistic smearing
(with the inner radius fixed at $6 \Rg$) and Comptonization smearing 
(with fixed $\tau = 0.7$ and $k T = 11$ keV) gives a rather bad fit, 
$\chi^2/d.o.f. = 150/126$.
The problem may be avoided by postulating that the hot skin is thicker
closer to the black hole (Nayakshin 2000), thus reducing the 
reflection amplitude from the inner regions. 

\subsubsection{Summary of untruncated disk models}

A fundamental problem is the high coronal dissipation required for
these magnetic flare models. The maximum which has so far been
produced in current (though admittedly incomplete) simulations is
$f_{cor}\sim 0.25$ (Miller \& Stone 2000), while the data imply
$f_{cor} > 0.5$. Apart from this the plasma outflow model can
reproduce the observations, though the complex soft excess is
unexplained. The hot skin model gives qualitative origin for the soft
excess as Compton scattering of the disk photons, but quantitatively
this does not give what is observed. Also the Comptonization in the 
hot skin gives an additional smearing of the iron line and the reprocessed 
spectrum, and
this smearing is much larger than that observed. However, if the model
is extended to allow radial stratification of the depth of the hot
skin then this may allow both the soft excess and line smearing to
match that observed.

\section{Conclusions}

Broad band 0.1--200 keV BeppoSAX data from Cyg X--1 in low/hard state
can be described by a complex soft component and a comptonized hard
component with the corresponding X-ray reprocessed spectrum. The soft
component can be decomposed into a low temperature ($k T \approx 0.1$
keV) blackbody emission, presumably from the accretion disk, and an
additional component which can be modelled as Comptonization of the
disk emission by a low temperature ($k T \sim 2$ keV) plasma.  
The hard comptonized
component corresponds to a spectrum with photon index $\sim 1.6$ with
a cutoff at the electron temperature of the corona, $\kT \sim 100$
keV.  The reflection spectrum consists of a weakly ionized, smeared
component.
Assuming that this broadening is due to 
relativistic smearing, this corresponds
to an inner disk radius of $\sim 70\,\Rg$ or $\sim 10\,\Rg$ 
(depending on the inclination of the system).  A non-smeared
component, due to reflection from the companion star and/or a flared outer 
disc, might also be present.  In this case we find that the inner radius
of the disk, as deduced by the reflection component, is smaller, $\sim
6-10$ $\Rg$.  

The spectrum is broadly compatible with the three major scenarios
proposed for emission in GBHCs, although all need some modification to 
fit the data. These are: (1) a hot ($\kT\sim 100$ keV),
optically thin(ish),  inner flow and a cold outer accretion disk, 
an accretion disk with active regions with either (2) mildly relativistic
outflow ($v \sim 0.3 c$) of the comptonizing plasma or (3) a hot ($k T \sim
10$ keV), ionized skin.

\acknowledgments
We thank B. Czerny and A. Zdziarski for useful discussions.
This work was supported by the Italian Space Agency (ASI), by the Ministero
della Ricerca Scientifica e Tecnologica (MURST) and in part by Polish KBN 
grants 2P03D01816 and 2P03D01718.


\clearpage

\section*{TABLES}

\begin{table}[h!]
\begin{center}
\caption{Results of the fit in the energy band 0.1--200 keV.  
Error bars correspond to $\Delta \chi^2 = 2.7$. Blackbody flux is 
in units of $L_{39}/D_{10}^2$, where $L_{39}$ is the luminosity 
in units of $10^{39}$ ergs/s and $D_{10}$ is the distance in units of 10 kpc.
{\it thComp\/} normalization is in photon cm$^{-2}$ s$^{-1}$ at 1 keV.
$f = \Omega/2\pi$ is the reflection amplitude ($f=1$ corresponds to the 
amplitude of reflection expected from a slab subtending a $2\pi$ solid
angle around an isotropic source)
and $\xi$ is the ionization parameter, $\xi=L_{\rm X}/n_{\rm e} r^2$.}
\label{tab1}
\begin{tabular}{l|c|c|c|c} 
\tableline \tableline
 Parameter    & Model 1 & Model 2  & Model 3 & Model 4 \\ \tableline
$N_{\rm H}$ $(\times 10^{22}$ cm$^{-2})$ & $0.70^{+0.12}_{-0.09}$ &
$0.69 \pm 0.11$ & $0.74 \pm 0.14$ & $0.85^{+0.13}_{-0.10}$ \\
$kT_{\rm BB}$ (keV)        & $0.13^{+0.27}_{-0.01}$ &
$0.130 \pm 0.012$ & $0.124 \pm 0.014$ & $0.112 \pm 0.010$ \\
F$_{\rm BB}$            & $0.049^{+0.075}_{-0.027}$ &
$0.045^{+0.062}_{-0.023}$ & $0.07^{+0.13}_{-0.05}$ & $0.16^{+0.26}_{-0.08}$\\
$kT_{\rm comptt}$ (keV)    & $2.6^{+4.9}_{-0.7}$ &
$3.0^{+1.8}_{-0.9}$ & $3^{+10}_{-1}$ & $2^{+10}_{-1}$ \\
$\tau_{\rm comptt}$        & $8.6^{+2.4}_{-8.6}$ &
$8.3^{+2.4}_{-5.3}$ & $7.5^{+2.2}_{-7.5}$ & $5.7^{+6.8}_{-5.7}$ \\
N$_{\rm comptt}$        & $0.79^{+0.76}_{-0.33}$ &
$0.65^{+0.59}_{-0.24}$ & $0.8^{+1.3}_{-0.7}$ & $4.0^{+3.6}_{-2.3}$ \\
Photon Index           & $1.614 \pm 0.013$ &
$1.619 \pm 0.015$ & $1.627 \pm 0.014$ & $1.634 \pm 0.013$ \\
$kT_{\rm e}$ (keV)           & $111 \pm 19$ &
$123^{+28}_{-20}$ & $138^{+41}_{-24}$ & $140^{+50}_{-35}$ \\
N$_{\rm thComp}$             & $1.086^{+0.047}_{-0.061}$ &
$1.079 \pm 0.050$ & $1.10^{+0.04}_{-0.13}$ & $1.205 \pm 0.040$ \\
f$_{\rm rel}$              & -- &
$0.321 \pm 0.033$ & $0.125^{+0.041}_{-0.060}$ & $0.185 \pm 0.040$ \\
Fe abund               & 1.0 (fixed) & 1.0 (fixed) & 1.0 (fixed) &
2.0 (fixed) \\
$\cos i$               & 0.6 (fixed) & 0.6 (fixed) & 0.6 (fixed) &
0.78 (fixed) \\
$\xi$                  & $12.9^{+6.1}_{-5.1}$ &
$15.4^{+6.7}_{-4.6}$ & $90^{+500}_{-50}$ & $35^{+25}_{-27}$ \\
$R_{\rm in}/\Rg$         & -- &
$130^{+99}_{-45}$ & $17^{+29}_{-7}$ & $6^{+5.5}$ \\
$f_{\rm narrow}$           & $0.288 \pm 0.027$ &
-- & $0.249^{+0.052}_{-0.066}$ & $0.12 \pm 0.03$ \\
$\chi^2$/d.o.f.        & 713/679 & 689/678 & 683/677 & 654/677  \\
\tableline
\end{tabular}
\end{center}
\end{table} 

\begin{table}[th]
\begin{center}
\caption{Results of the fit in the 0.1--200 keV energy range with one
reflection component for different values of the inclination angle and 
of the iron abundance: $\chi^2 ~(R_{\rm in})$, where $R_{\rm in}$ is in
units of $\Rg$, and the degrees of freedom are 678. Typical errors on
$R_{\rm in}$ are as in Table 1.}
\label{tab2}
\begin{tabular}{|l|c c c c c|}  \tableline
Fe abund & \multicolumn{5}{|c|}{Inclination ($\cos i$)} \\ 
       & 0.87       & 0.78       & 0.71       &  0.60     &   0.50     \\  
\tableline
3        & 675 (8.3)  & 685 (6.0)  & 701 (42.5) & 704 (34.7) & 700 (47.7) \\
2        & 669 (9.85) & 676 (40.2) & 679 (52.4) & 687 (57.3) & 685 (76.6) \\
1.5      & 673 (15.6) & 670 (69.0) & 674 (79.7) & 679 (84.5) & 679 (109)  \\
1        & 686 (49.4) & 687 (91.4) & 688 (117)  & 689 (130)  & 690 (167)  \\
\tableline
\end{tabular}
\end{center}
\end{table} 

\begin{table}[h!]
\begin{center}
\caption{Results of the fit, $\chi^2$ ($R_{\rm in}$), in the 0.1--200 keV 
energy range with two reflection components for different 
values of the inclination angle and of the iron abundance. $d.o.f. = 677$.
Typical errors on $R_{\rm in}$ are as in Table 1.
There is a clear trend in the $\chi^2$ values to be smaller at super-Solar 
iron abundance and for $\cos i \sim 0.8$ ($i=35^\circ - 40^\circ$).}
\label{tab3}
\begin{tabular}{|l|c c c c c|}  \tableline
Fe abund & \multicolumn{5}{|c|}{Inclination ($\cos i$)} \\ 
       & 0.87      &  0.78     & 0.71       &  0.60      & 0.50  \\  
\tableline
3        & 676 (6.0) & 659 (6.0) & 663 (6.0)  & 675 (12.0) & 677 (15.0)   \\
2        & 664 (9.5) & 654 (6.0) & 658 (10.2) & 663 (14.3) & 663 (17.5)  \\
1.5      & 674 (6.0) & 658 (6.0) & 664 (6.0)  & 668 (16.8) & 668 (20.9)  \\
1        & 685 (6.0) & 667 (6.0) & 674 (6.0)  & 683 (17.3) & 686 (26.4) \\ 
\tableline
\end{tabular}
\end{center}
\end{table} 

\clearpage
 
\section*{FIGURES}

\begin{figure}[h]
\centerline
{\psfig
{figure=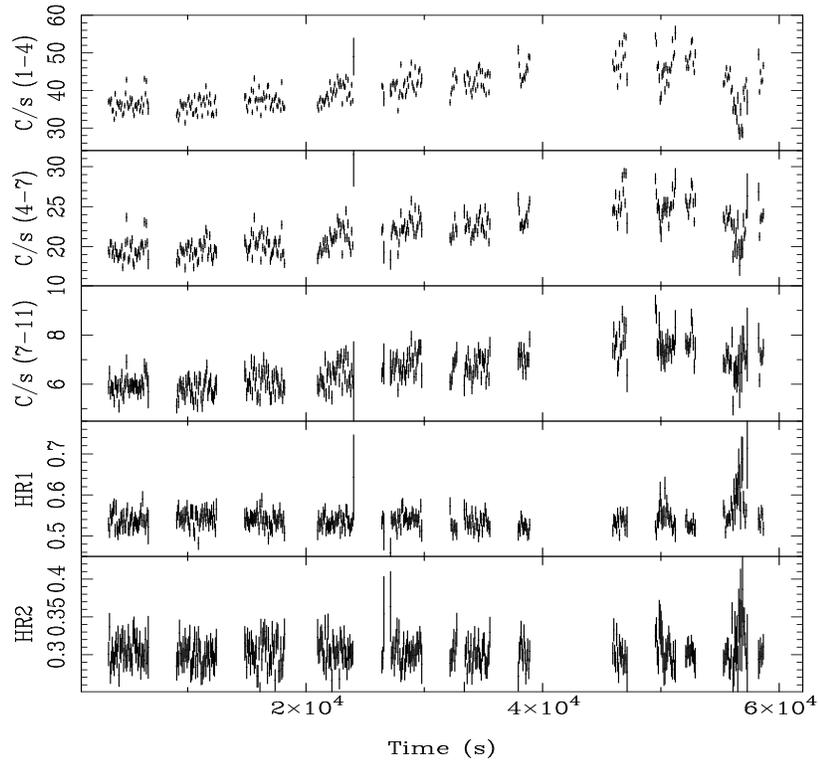,height=12.0cm,width=12.0cm}}
\caption{MECS light curves of Cyg X--1 in three energy bands: 
1--4 keV (upper panel), 4--7 keV (top middle panel), and 7--11 keV (middle 
panel), and the corresponding hardness ratios: HR1 = 4--7 keV/1--4 keV 
(bottom middle panel) and HR2 = 7--11 keV/4--7 keV (lower panel).}
\label{fig1}
\end{figure}

\begin{figure}[h]
\centerline
{\psfig
{figure=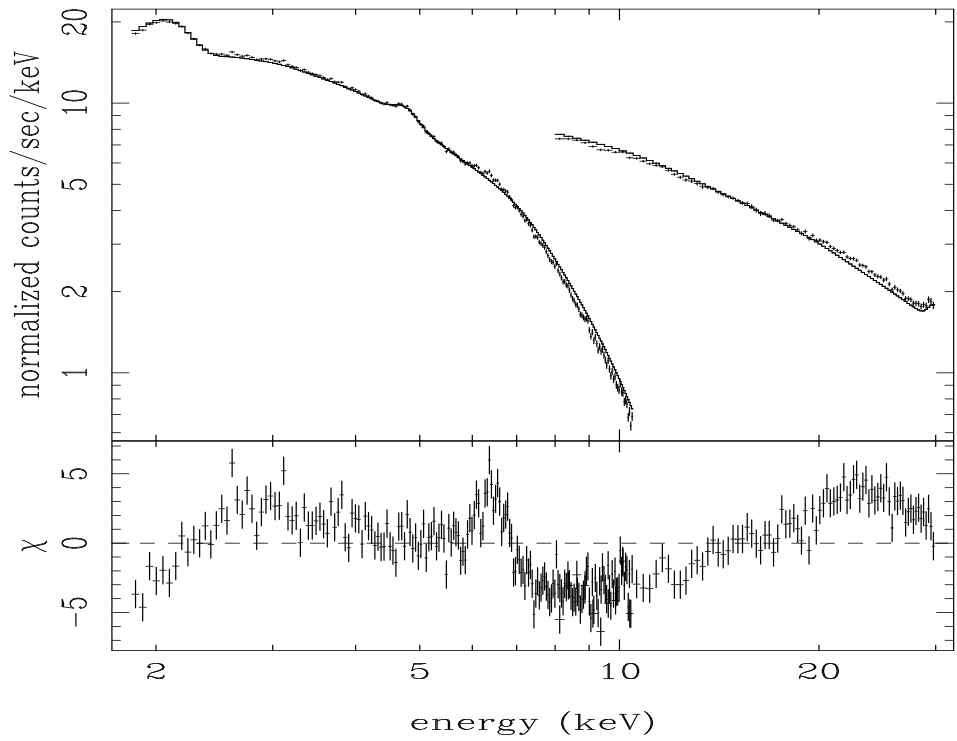,height=12.0cm,width=14.0cm}}
\caption{Result of a fit of the 1.8--30 keV spectrum of Cyg X--1 
with a simple photoelectric absorbed power law (the $\chi^2/d.o.f.$
was 2008/274).  Upper panel: MECS and HPGSPC data and model; lower
panel: residuals (in unit of $\sigma$) with respect to the power law. }
\label{fig2}
\end{figure}

\begin{figure}[h]
\centerline
{\psfig
{figure=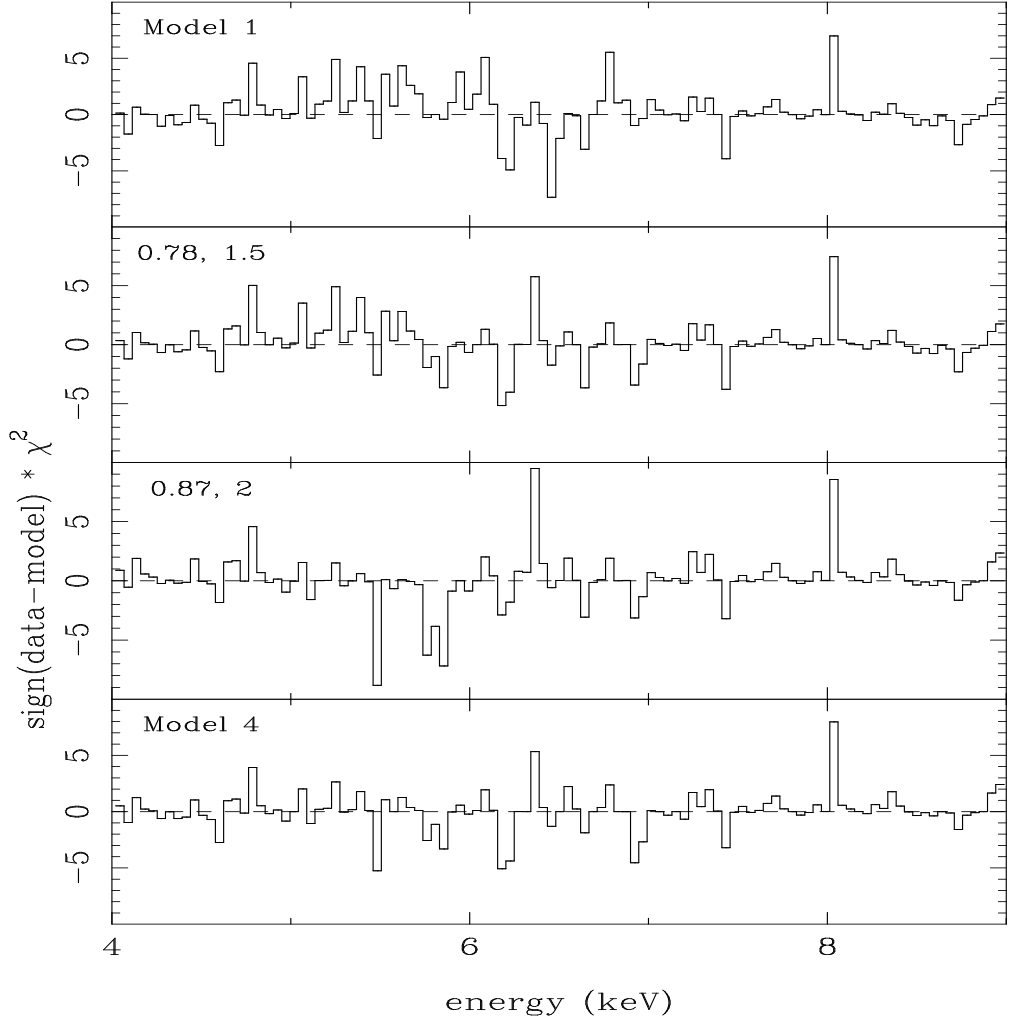,height=14.0cm,width=14.0cm}}
\caption{Contributions to the $\chi^2$ as a function of the energy 
in the iron line range for various fit models.  Upper panel:
model 1 of Table 1 with a not relativistically smeared reflection.
Top middle panel: model corresponding to $\cos i = 0.78$
and [Fe] = 1.5. Bottom middle panel: model corresponding
to $\cos i = 0.87$ and [Fe] = 2.  Lower panel:  model consisting of
two reflection components for $\cos i = 0.78$ and [Fe] = 2 (model 4 in
Table 1). }
\label{fig3}
\end{figure}

\begin{figure}[h]
\centerline
{\psfig
{figure=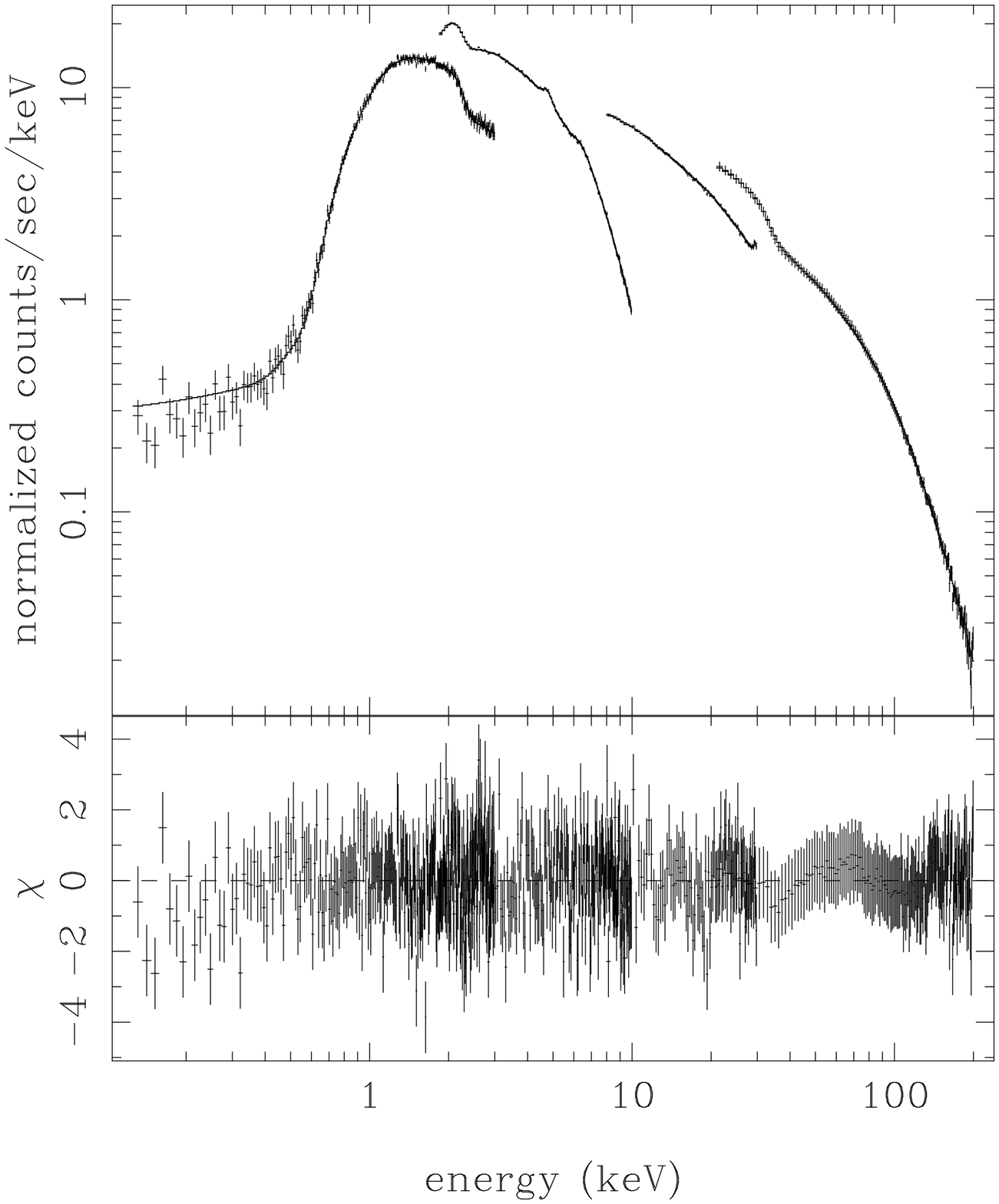,height=12.0cm,width=14.0cm}}
Fig. 4a. -- Cyg X--1 data fit using model 4 of Table 1 
(upper panel) and the residuals (in unit of $\sigma$) with respect 
to the model (lower panel). 
\label{fig4a}
\end{figure}

\begin{figure}[h]
\centerline
{\psfig
{figure=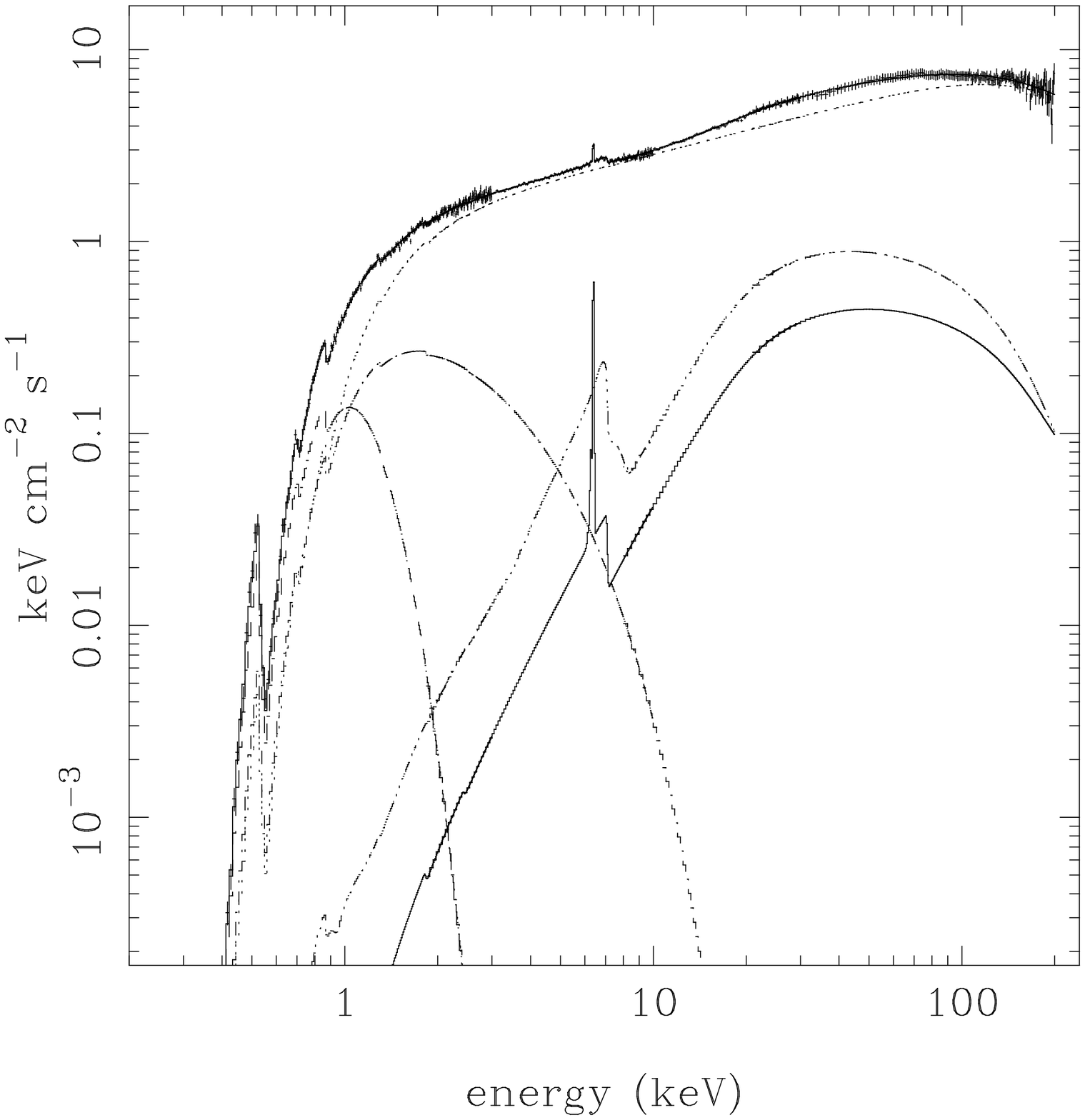,height=14.0cm,width=14.0cm}}
\vskip -1.7cm
Fig. 4b. -- Cyg X--1 unfolded spectrum predicted by model 4 (Table 1).  
The single components of the model are also shown, namely
the soft component, described by the blackbody at $\sim 0.1$ keV 
(dashed line) and the low temperature ($\sim 2$ keV) comptonized component
(dot-dashed line), the hard 
comptonized emission from a hot ($\sim 140$ keV) corona (dotted line),
and the corresponding narrow and smeared reprocessed components (solid
and dot-dot-dot-dashed line respectively)
containing the iron K$_{\alpha}$ line and edge, and the reflected continuum. 
\label{fig4b}
\end{figure}

\end{document}